\documentclass{aa}
\usepackage{graphics}
\newcommand{\lya}{\hbox{Ly$\alpha$}}

\newcommand{\etal}{\hbox{et al.\ }}
\begin{document}
   \thesaurus{01     % A&A Section 1. Letters
              (11.01.2;  % Galaxies: active
               11.03.1;  % Galaxies: clusters: general
               11.05.2;  % Galaxies: evolution
               12.03.3;  % Cosmology: observations
               12.05.1; % Cosmology: early Universe
               12.12.1)} % Cosmology: Large-scale structure
\title{A search for clusters at high redshift}
 \subtitle{II. A proto cluster around a radio galaxy at $z=2.16$}

\author{L. Pentericci \inst{1,2} \and
        J.D. Kurk \inst{2} \and H.J.A R\"ottgering \inst{2} \and
        G.K. Miley \inst{2} \and W. van Breugel \inst{3}  
        \and C.L. Carilli \inst{4} \and H. Ford \inst{5}
        \and T. Heckman\inst{5} \and P. McCarthy \inst{6} 
        \and A. Moorwood \inst{7} }
%          \and
%          C. Ptolemy\inst{2}\fnmsep\thanks{Just to show the usage
%          of the elements in the author field}
\offprints{L. Pentericci (laura@mpia-hd.mpg.de)}

\institute{Max-Planck-Institut f\"ur Astronomie, K\"onigstuhl 17,
              D-69117, Heidelberg, Germany
         \and
              Sterrewacht Leiden, P.O. Box 9513, 2300 RA, Leiden,
              The Netherlands
         \and
              Lawrence Livermore National Laboratory, P.O. Box 808,
              Livermore CA, 94550, USA
         \and
              NRAO, P.O. Box 0, Socorro NM, 87801, USA
        \and
              Dept. of Physics \& Astronomy, The Johns Hopkins University,
              3400 North Charles Street, Baltimore MD, 21218-2686, USA
          \and
              The Observatories of the Carnegie Institution of Washington,
              813 Santa Barbara Street, Pasadena CA, 91101, USA
        \and
              European Southern Observatory, Karl-Schwarzschild-Str. 2,
              D-85748, Garching bei M\"unchen, Germany
           }
%         \and
%             University of Alexandria, Department of Geography\\
%             email: c.ptolemy@hipparch.uheaven.space
%             \thanks{The university of heaven temporarily does not
%                     accept e-mails}
\date{Received; accepted }
\maketitle

\begin{abstract}
VLT\footnote{Based on observations carried out at the European

Southern Observatory, Paranal, Chile, programme 64.O-0134.} 
spectroscopic observations of \lya\--excess objects
 in the field of the clumpy
radio galaxy 1138--262 at $z=2.16$ have led to the discovery of
14 galaxies and one QSO at approximately the same distance as the 
 radio galaxy. All galaxies have 
redshifts in the range 2.16$\pm$0.02, centered around the redshift of the
radio galaxy, and are within a projected
physical distance of 1.5 Mpc from it. 
The velocity
distribution suggests that there are two galaxy subgroups having
velocity dispersions of $\sim 500$ km s$^{-1}$ and $\sim 300$ km
s$^{-1}$ and a relative velocity of 1800 km
s$^{-1}$. If these are virialized structures, the estimated
dynamical masses for the subgroups are $\sim$9 and $\sim$3 $\times$
10$^{13}$~M$_{\odot}$ respectively, implying a total mass for the
structure of more than 10$^{14}$~M$_{\odot}$. The new observations,
together with previous results, 
suggest that the structure of galaxies around 
1138--262 is likely to be a forming cluster.
% and (ii)
%that active galaxies are important for probing the formation of
%structure in the early Universe.

%We also detect Ly$\alpha$ emission from several components of the
%radio galaxy, which show a similar velocity dispersion.

\end{abstract}
%
%________________________________________________________________
\section{Introduction} One of the most intriguing questions in
modern astrophysics concerns the formation of large scale
structure in the early Universe (e.g.\ Bahcall \etal 1997). Although various scenarios have been
developed within the context of modern cosmological models,
 the epoch and mechanism of the formation of galaxy clusters 
are still open questions.
\\\indent
Despite several tentative identifications of clusters and groups at
high redshift (e.g. Keel \etal  
1999; Campos et al. 1999; Pascarelle \etal 1998), 
there is yet no solid identification of collapsed
clusters or groups at redshifts above 1.5. Recently Steidel \etal
(2000) reported the discovery of a significant overdensity of
galaxies at $z \sim 3.1$ and pointed out its possible association
with a quasar.
%Here we report the detection of an even more
%substantial overdensity of galaxies around a  $z=2.2$ radio
%galaxy, that is an X-ray emitter and has several of the properties
%expected from a forming massive galaxy at the center of a
%cluster.
\\
During the last few years evidence that distant radio
galaxies may be  located in over-dense environments has
mounted (e.g. R\"ottgering \etal 1999). Because such
objects can be detected out to $z > 5$, we have begun a programme
to survey their surroundings for galaxy overdensities.
As a good candidate for a pilot project we selected the powerful
radio galaxy 1138--262\ from a compendium of more than 150 $z
> 2$ known radio galaxies, for several reasons, including
(i) its extremely clumpy optical morphology and large size that
resembles simulations of forming massive galaxies (Pentericci \etal 1998); 
(ii) its extreme Faraday rotation and distorted radio morphology 
indicating a dense magnetized surrounding gas and (iii) its detection of 
possibly extended X-ray emission (Carilli
\etal 1998).
\\\indent
Deep VLT narrow and broad band imaging of 1138--262, carried
out during April 1999 (Kurk \etal 2000, hereafter Paper I), 
revealed 50 candidate Ly$\alpha$ emitters in
a $7' \times 7'$ field around the radio source. The
derived luminosity function indicated a possible overdensity of
galaxies in this field, expecially considering 
that a large fraction of the
starburst galaxies and the more passive elliptical galaxies
 would not show \lya\ emission. 
Therefore these observations strongly suggested the presence of a
forming  cluster around 1138--262.
Here we report on follow up spectroscopic
observations of these cluster candidates.  
\footnote{
Throughout this paper we assume a Hubble constant of $H_0 =50$ km s$^{-1}$ 
Mpc$^{-1}$ and a deceleration parameter of $q_0 =0.5$}

%__________________________________________________________________
\section{Selection, observations and data reduction}
The observations were carried out on March 1, 2 and April 7, 2000 with 
the 8.2m VLT Antu telescope
(UT1), using the FORS1 imaging spectrograph in the multi-object
spectroscopic (MOS) mode with standard resolution, giving  
a pixel scale of 0.2$''$ and a field of view of 6.8$'
\times$ 6.8$'$. The first two nights  were
photometric, the third was not. 
The spectra were obtained with the 600B grism, using 
a wavelength range from 3450 to 5900 \AA, depending on the
position of the slit in the set--up and with a dispersion of 1.2 \AA\
pixel$^{-1}$.

Three different set--ups were used, each with 19 slits. We
have chosen a slit size of 1$''$, 
which gives a resolution of $\sim 5$ \AA\ corresponding
to about 400 km s$^{-1}$ at $z=2.16$. The MOS set--up
constrains the choice of targets. 
Therefore,  
to allow a wider choice of targets, we included in our list 
of  candidate \lya\ emitters not only the 50 objects selected in Paper I,
that had equivalent width (EQW) $\ge 65$ \AA\ (corresponding 
to a rest-frame EQW of 20
\AA) and a narrow band flux density 
$\ge$ 2 $\times\ 10^{-19}$ ergs s$^{-1}$ cm$^{-2}$ \AA$^{-1}$,
 but also those with fluxes below this limit.
The total number of possible targets 
was then 75 instead of 50.

Of the 57 slits available, we
placed 2 on different components of the central radio
galaxy, and 46 on candidate \lya\ emitters (3 of which were
observed twice). Of the remaining 9 slits, 3 (one in each
set--up) were placed on relatively bright stars to check that the
slits were positioned properly, and 6 were
placed on additional objects, with EQW
just below the cutoff adopted in Paper I.
\\
The observations were divided into
multiple one hour exposures to facilitate
removal of cosmic-ray effects from the final image. The
first set--up was observed for a total of 6 hours, the second
for 5.5 hours, and the third for 4 hours.
All observations were carried out at airmass less than 1.5.
%\vspace{0.3cm}

The data were reduced using standard IRAF procedures: after bias
subtraction and flat field removal, for each slit the background
was subtracted from the data avoiding regions where the spectrum
was visible. A relative flux calibration was obtained from a
longslit observation with the 600B grating of the
spectrophotometric standard star GD108 (Oke 1990). Wavelength
calibration was obtained from exposures of He and HgCd arc-lamps 
taken on the afternoon before the
observations. The accuracy of the wavelength
calibration was better than 0.2\AA.  One--dimensional 
spectra were extracted for all objects, using aperture
sizes which included all the emission.

% THIS TABLE COULD WELL BE IN SMALLER FONT??

\begin{table*}
\begin{center}
\caption{\rm Emitter characteristics}
\begin{tabular}{r c c c  c c c } \hline \hline 
Object  & Position &  $z$  & Flux  &  EQW & FWHM & B-mag\\
 \hline 
& R.A. (2000) \hskip0.2cm Decl. (2000) &   &10$^{-17}$ ergs s$^{-1}$cm$^{-2}$& \AA &
 km s$^{-1}$ \\
\hline 
7   & 11 40 37.14  \hskip0.2cm-26 32 08.2  &2.143&5.6 &63     &510 & 25.4\\
387 & 11 40 55.31  \hskip0.2cm-26 30 43.5  &2.139&1.7 &22     &370 & 28.1\\
419 & 11 40 55.35  \hskip0.2cm-26 30 36.9  &2.141&2.1 &$\ge$136&940& 28.0\\
724 & 11 40 57.50  \hskip0.2cm-26 29 39.4  &2.164&2.5 &16     &200 &--   \\
752 & 11 40 58.20  \hskip0.2cm-26 29 34.1  &2.170&5.5 &122    &540 &--   \\
759 & 11 40 46.89  \hskip0.2cm-26 29 32.2  &2.145&1.4 &$\ge$88&--  &--   \\
833 & 11 40 46.21  \hskip0.2cm-26 29 03.2  &2.155&4.0 &$\ge$60&480 &--   \\
856 & 11 40 49.41  \hskip0.2cm-26 29 09.4  &2.166&4.6 &35     &220 & 24.8\\
1189& 11 40 57.11  \hskip0.2cm-26 28 11.0  &2.165&3.5 &59     &490 & 26.3\\
1240& 11 40 59.16  \hskip0.2cm-26 27 55.3  &2.147&3.3 &14     &790 & 25.1\\
1405& 11 40 44.45  \hskip0.2cm-26 27 43.4  &2.164&3.1 &155    &470 & 26.2\\
1518& 11 40 36.88  \hskip0.2cm-26 28 03.3  &2.161&8.8 &72     &830 & 25.0\\
1557& 11 40 59.07  \hskip0.2cm-26 28 10.6  &2.147&2.7 &13     &260 & 25.4\\
1612& 11 40 55.28  \hskip0.2cm-26 28 24.3  &2.163&3.9 &$\ge$30&490 & 27.4\\
QSO1687&11 40 39.76  \hskip0.2cm-26 28 45.3&2.183$^{\it a}$&56&164&5800 &  24.8\\

\hline\hline
\end{tabular}
\end{center}
$^{\it a}$ Uncertain due to the presence of absorption.
\end{table*}

%__________________________________________________________________

\begin{figure}
 \resizebox{9cm}{!}{\includegraphics{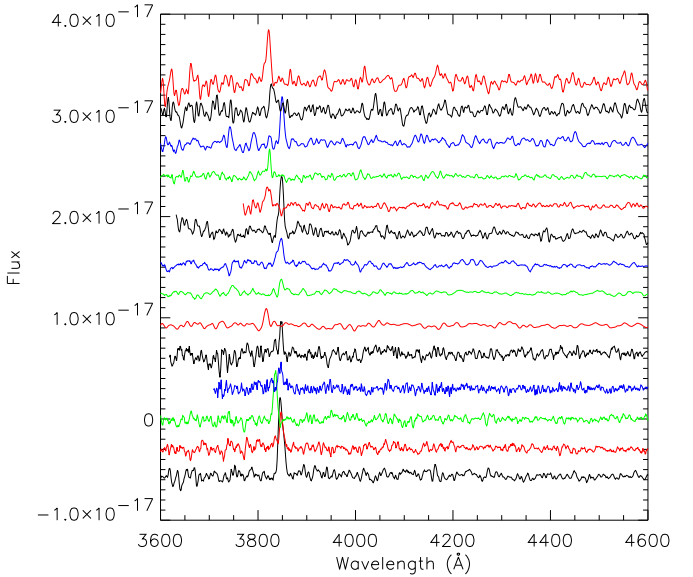}}
 \caption{The emission spectra of the 14 confirmed galaxies 
at redshift $\sim 2.16$. The flux is in ergs s$^{-1}$ cm$^{-2}$ A$^{-1}$
and for clarity each spectrum is offset, relative to the ordinate, by  
multiples of 3 $\times$ 10$^{-18}$ ergs s$^{-1}$ cm$^{-2}$ A$^{-1}$.}
\end{figure}
\section{Results}

For 15 candidates we obtained a clear detection of
 an emission line, which we identify as \lya\ at
redshift  of 2.16$\pm$0.02. Alternative line
identifications 
such as [\ion{O}{II}]$\lambda$3727 from $z \sim 0.02$ objects are
unlikely, since the H$\beta$ and [\ion{O}{III}]5007 lines should
then be visible in the spectra. Furthermore, the probability to
observe a foreground galaxy is very small, since the differential
volume element at $z \sim 0.02$ is 370 times smaller than that at
redshift 2.16.
One of the 15 emitters is an AGN.
We also detect emission from several components of the central
radio galaxy  1138--262, which we shall discuss
 in a companion paper (Kurk \etal \emph{in preparation}). 
\\
Besides the 15 \lya\ emitters, an additional 13
objects show (mostly faint) continuum emission, but none have detectable lines
therefore their redshifts remain undetermined.
Five  of these 13 objects 
show a decrement in continuum around the wavelength which would correspond
to a Lyman break at a redshift $\sim 2$, so they could also
 be high redshift galaxies. Alternatively this spectral shape could be 
attributed to evolved stellar populations in faint low redshift galaxies. 
Note that in general the objects observed do not have strong continuum
emission, because of our selection criteria.

The success rate of our selection criteria is 70\%
for prime candidates (i.e. having an observed EQW $> 195$ \AA),
40\% for secondary candidates
(having EQW $> 130$ \AA), 19\% for tertiary targets (EQW $> 65$ \AA), 
and 17\% for the rest.
The undetected objects with large EQW 
all have low expected \lya\ flux. Therefore the
non--detections are probably due to a lack of signal to noise.

Spectra of the newly discovered galaxies are shown in Fig.\ 1 and
that of the AGN in Fig.\ 2. In Table 1 we report the
positions, the redshift calculated from the peak position of the
\lya\ emission, the total flux in ergs s$^{-1}$ cm$^{-2}$, the
rest frame EQW, the restframe deconvolved FWHM 
of the \lya\ line in km s$^{-1}$,
determined by fitting the emission with a Gaussian function.
Most of the emission lines have, to a first approximation, symmetric shapes,
although in some cases the lines show structures and in one object 
a velocity gradient. We will consider this aspect further in follow-up work.
In Table 1 we also report 
the B-band magnitude determined from the broad band observations
of Paper I. 
%Typical sizes of the emitters are 1 -- 2$''$, which
%means that in most cases by using a 1$''$ slit we 
%do not measure all of the flux.  For this reason, and
%to ensure measurements with
%the same atmospheric conditions, the total fluxes 
%given in Table I are derived from the narrow band
%observations (Paper I).
\begin{figure}
 \resizebox{8.5cm}{!}{\includegraphics{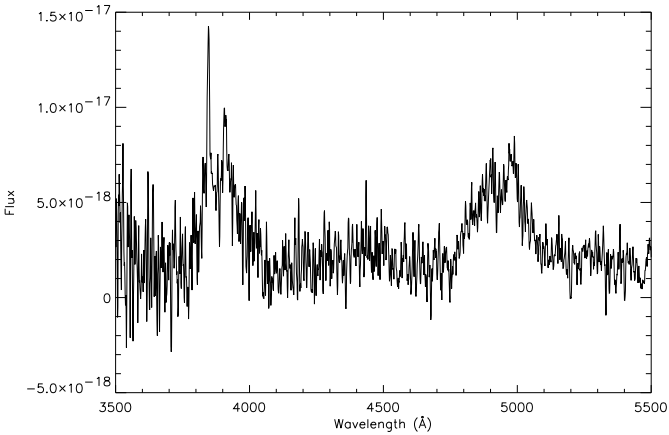}}
 \caption{The emission spectrum of QSO 1687, showing broad \lya, \ion{Si}{IV} and \ion{C}{IV} emission lines. \lya\ and \ion{C}{IV} are partly absorbed.}
\end{figure}
\\
The mean rest frame EQW for our sample of
galaxies is 60 \AA\ and the distribution is nearly uniform from 15 to 150 \AA. 
There are several objects with EQW in excess of 100 \AA.
This is significantly different from what is obtained by Steidel \etal (2000)
for a large sample of galaxies at redshift around 3: they 
find no galaxies with EQW larger than 100 \AA. 
While our results are most probably due to our observational
strategy, that emphasizes large EQW objects,
 it is also possible that some of our targets contain an AGN
component, since photoionization from hot stars is unlikely to
produce rest frame EQW much higher than 100 \AA\ (e.g.\ Charlot \&
Fall 1993).
\\\indent
The QSO shows \lya, \ion{Si}{IV} and \ion{C}{IV} emission
lines, and the \lya\ has a FWHM of 5800 km s$^{-1}$. The \lya\ and \ion{C}{IV}
lines have a double--peaked profile that is
due to absorption by neutral gas. The fact that also \ion{C}{IV}
is absorbed indicates that the gas is metal enriched (e.g.\
Binette \etal 2000). The \lya/\ion{C}{IV} line ratio appears
depressed compared to normal values for QSOs (\lya/\ion{C}{IV}
$\sim 2.5$ from composite spectra of QSO BLRs, Osterbrock 1989).
This could be caused by the presence of dust.

%__________________________________________________________________

\section{Discussion}
In Fig.\ 3 we show the redshift distribution of the
15 newly discovered galaxies, together with a Gaussian
curve representing the approximate
normalized sensitivity of the narrow band filter, used at the
beginning of the project to select the candidate emitters (Paper
I). The velocities of the galaxies with confirmed redshifts are
not distributed according to the selection function, but appear
to be clustered around the redshift of
the central radio source ($z = 2.156$).
\begin{figure}
\resizebox{8.5cm}{!}{\includegraphics{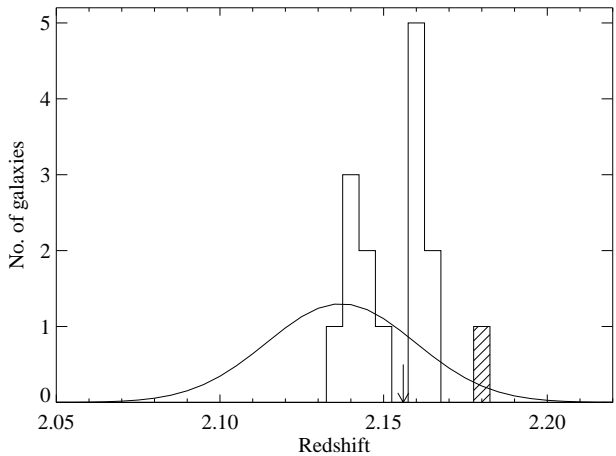}} \caption{The
redshift distribution of the 15 emitters. The hatched bin contains
the newly discovered QSO and the arrow denotes the redshift of
the  radio galaxy. The Gaussian curve represents the
approximate response of the narrow band
filter used to select the candidates.}
\end{figure}

It is instructive to compare the total comoving volume density of this
structure of galaxies to that of 
the spike at redshift 3.09, found by
Steidel \etal (2000), which consists of 24 galaxies and 
represents an overdensity of
a factor $\sim 6$ compared to the field. Note that only 12
of these 24 galaxies have been subsequently 
also found using NB selection 
criteria similar to ours: the
remaining ones have weak or absent Ly$\alpha$ emission or
 Ly$\alpha$ deficits. \\
For a meaningful comparison, we adopt the same intrinsic 
luminosity cut as Steidel et al. (corresponding to 
their NB mangnitude limit of NB $\le$ 25, in
AB magnitudes).
This brings down the number of emitters we would have detected in 
our field to 4 (including the central radio galaxy).
We then multiply this number by a factor two to compensate for the 
galaxies without \lya\ emission. 
The comoving volume for our 
field size (8.2 Mpc$^2$) and redshift range ($2.139 \le z \le 2.170$)
is 3830 Mpc$^3$, resulting in a volume density of 2.1 $\times 10^{-3}$ 
objects Mpc$^{-3}$.
This is approximately equal to the density of the Steidel et al. spike, with
24 galaxies in a total comoving volume of 11040 Mpc$^3$, resulting 
in a density of 2.2 $\times 10^{-3}$ objects Mpc$^{-3}$.

The following obvious question is whether the overdensity of galaxies around
1138--262 is due to a structure that is, or will become, a gravitationally 
bound cluster. To investigate this matter further we discuss some
additional properties of the data. 
The total velocity dispersion of the structure is 1200 $\pm$160 km s$^{-1}$
(1000$\pm$140 km s$^{-1}$ excluding the QSO), calculated as the 
standard deviation from the mean,
i.e. assuming an underlying Gaussian
velocity distribution. This would imply 
a mass larger than 5 (3.5) $\times$ 10$^{14} M_{\odot}$, assuming a 
radius of 1.5 Mpc. However, the velocity
distribution of the galaxies (Fig.\ 3)  clearly differs 
from a Gaussian. The galaxies seem to cluster 
in two subgroups with seven members each, centered at median redshifts of
2.145$\pm0.002$ and 2.164$\pm0.002$, with the QSO as an outlier. 
Using the gapper sigma,
suggested by Beers \etal (1990) as a better scale estimator for
very small samples of objects, we obtain a velocity dispersion
of 520$\pm 140$ km s$^{-1}$ and 280$\pm 70$ km s$^{-1}$
respectively. The implied total masses of each subgroup
separately are considerably lower, $\sim 9 \times 10^{13} M_{\odot}$ 
and $\sim 3 \times 10^{13} M_{\odot}$. The detection of a larger number
of galaxies is needed to establish
the internal kinematics of these structures 
so that more definitive statements can
be made about the virialization and mass distribution.
\\\indent
To further investigate possible spatial
clustering of the galaxies we computed the angular
two--point correlation function. The
 subgroups are too small to obtain any
significant statistical results, but using the Landy-Szalay
estimator (Landy \& Szalay \cite{lan93}) for the 15 galaxies and
1138--262, we detect a signal at distances of $\sim 25''$ due to
the occurrence of 6 close pairs in our sample and at $\sim 150''$ with
significance of 99.7\% and 99.8\% respectively.
We also used the estimator introduced by Phillips (1985) 
to measure the correlation between the
positions of observed objects and the position of a known
object (in this case the radio galaxy), but 
the distribution of emitters around
1138--262 was not found to be significantly different
from a random one.
\\
Although we find no significant spatial segregation 
 between the two  subgroups, this may be masked
by (i) the small number of galaxies so far detected, (ii) bias
introduced by constraints in the slit positions simultaneously
accessible to the FORS MOS and (iii) our identification of all
clumps within the 150--kpc sized \lya\ halo surrounding 1138--262
as components of the central radio galaxy.
\\
We note that the few high redshift
clusters known show similar substructures and do not appear
 concentrated in the sky.
Examples include the cluster around 3C234
(e.g. Dickinson 1997) at redshift $\sim$ 1.1; the
concentration of red objects at $z \sim$ 1.3 found by Liu \etal (2000),
 and the structure found by Rosati
\etal at redshift $\sim$ 1.27, consisting of two collapsed
systems (Rosati \etal 1999; Stanford \etal 1997).
Similarly, Steidel et al. (2000) find that the distribution of the
local density of galaxies for their z$=$3.09 spike,
has 3 significant density peaks. 
\\
Cluster substructures are not only frequently observed, but
they are also consistent with the predictions of hierarchical
models of galaxy and structure formation.
As Governato \etal (1998) argue, such concentrations of galaxies
at high redshift are  probably the progenitors of local rich
clusters. Given the small velocity separation, the velocity
subgroups found in 1138--262 are likely to merge, evolving into the larger
structures seen in the local Universe.

%__________________________________________________________________

\section{Conclusions}
We have confirmed the existence of a substantial overdensity of galaxies 
within 1.5 Mpc of the high redshift radio galaxy 1138--262.
We have found at least 15 galaxies at approximately the same distance as the
radio source which seem to consist of  two subgroups. The
new results taken together with previous work show that 1138--262
at $z = 2.16$ has many ingredients that might be expected from a
forming cluster, namely a substantial galaxy overdensity, 
a central galaxy resembling a forming massive central cluster galaxy
and the presence of hot X-ray--emitting gas. 
These results also confirm that radio galaxies can be used to pinpoint
regions of galaxies overdensity at high redshifts and to probe the formation
of large--scale structures in the early universe.
%__________________________________________________________________
\begin{acknowledgements}

The work by WvB at IGPP/LLNL was performed under the auspices of the
US Department of Energy under contract W-7405-ENG-48. 
\end{acknowledgements}

%__________________________________________________________________

\end{document}